\documentclass{article}
\usepackage[T1]{fontenc} 
\usepackage[utf8]{inputenc} 
\usepackage{ismir,amsmath,cite,url}
\usepackage{amssymb,stmaryrd}
\usepackage{graphicx}
\usepackage{color}
\usepackage{xspace}
\usepackage{microtype}
\usepackage{multirow}
\usepackage{booktabs}
\usepackage{tabularx}
\newcolumntype{Y}{>{\centering\arraybackslash}X}
\usepackage{caption}
\usepackage{subcaption}  
\usepackage{listings}  
\usepackage{siunitx}  
\usepackage{dsfont}  
\usepackage{todonotes}
\usepackage{pdfpages}

\newcommand{\eg}{\emph{e.g.}\xspace}
\newcommand{\ie}{\emph{i.e.}\xspace}

\title{Improved symbolic drum style classification with grammar-based hierarchical representations}

\multauthor
{Léo Géré$^1$ \hspace{1cm} Nicolas Audebert$^{1,2}$ \hspace{1cm} Philippe Rigaux$^1$} { 
  $^1$ Conservatoire national des arts et métiers, CEDRIC, F-75141 Paris, France\\
  $^2$ Univ. Gustave Eiffel, ENSG, IGN, LASTIG, F-94160 Saint-Mandé, France\\
{\tt\small leo.gere@lecnam.net, nicolas.audebert@ign.fr, philippe.rigaux@cnam.fr}
}

\def\authorname{L. Géré, N. Audebert, and P. Rigaux}

\usepackage[bookmarks=false,pdfauthor={\authorname},pdfsubject={\papersubject},hidelinks]{hyperref}
\usepackage[capitalize,noabbrev]{cleveref}

\begin{document}

\maketitle

\begin{abstract}
Deep learning models have become a critical tool for analysis and classification of musical data. These models operate either on the audio signal, \eg waveform or spectrogram, or on a symbolic representation, such as MIDI.
In the latter, musical information is often reduced to basic features, \ie  durations, pitches and velocities.
Most existing works then rely on generic tokenization strategies from classical natural language processing, or matrix representations, \eg piano roll.
In this work, we evaluate how enriched representations of symbolic data can impact deep models, \ie Transformers and RNN, for music style classification.
In particular, we examine representations that explicitly incorporate  musical information \emph{implicitly} present in MIDI-like encodings, such
as rhythmic organization, and show that they outperform generic tokenization strategies. 
We introduce a new tree-based representation of MIDI data built upon a context-free musical grammar.
We show that this grammar representation accurately encodes high-level rhythmic information and outperforms existing encodings on the GrooveMIDI Dataset for drumming style classification, while being more compact and parameter-efficient.
\end{abstract}

\section{Introduction}\label{sec:introduction}

In the last few years, machine learning (ML) has significantly changed how the Music Information Retrieval (MIR) community deals with tasks such as style and composer classification, music generation, pitch and rhythm detection, etc.
Yet, training deep learning models on music raises the question of the representation of this data.
Depending on the input format (audio, MIDI, musical score\dots), different representations, \ie different \emph{encodings}, are possible. Each encoding has advantages and drawbacks: some representations, \eg waveforms, focus on raw low-level acoustic features, while others, \eg sheet music, encode high-level abstract semantics of the musical language.

While deep neural networks had great success on audio signal, \ie waveforms and spectrograms, machine learning for symbolic MIDI remains understudied. In this work, we seek to build effective representations of symbolic music, with a focus on recorded MIDI performances.
Multiple possible representations of MIDI music coexist in the literature.
Most of them contains only low-level information, such as the timing of the onset and the offset of each note and their velocity.
This is due to practical constraints: typical MIDI recordings usually do not contain any information about tonality, tempo, time-signature or rhythm.
Hence, a model trained on such MIDI samples typically needs to allocate a part of its weights to extract these relevant high-level features from the data.
Building better representations of MIDI data to encode semantic musical information could therefore be beneficial to the training of deep models and their efficiency, as they could directly focus on using these features rather than extracting them from the data first.

Music classification has been a task of choice for MIDI performances. Preliminary works from \cite{cataltepe2007music} in 2007 encoded MIDI as strings and used Kolmogorov complexity to compare music pieces. \cite{mckay2006jsymbolic} introduced jSymbolic, a library to extract high level features from MIDI files, such as pitch histograms, a line of work extended by \texttt{music21} \cite{cuthbert2011feature} and \texttt{musif} \cite{simonetta2023optimizing}. As new MIDI datasets have been introduced for composer \cite{kong2020large} and style classification \cite{groove2019}, efforts have been made to evaluate how MIDI representations affect deep models. \cite{miditok2021} introduced \mbox{{MidiTok}}, a tokenization framework to encode MIDI files as a sequence of tokens, suitable for Transformers and Recurrent Neural Networks (RNN). More recently, \cite{zhangSymbolicMusicRepresentations2023b} compared different neural architectures for various MIDI encodings: Convolutional Neural Networks (CNN) trained on Piano rolls, Transformers trained on sequences of tokens, and Graph Neural Networks (GNN) trained on graphs extracted from MIDI files.

In this line of work, we aim to design a representation of MIDI files that is both efficient and discriminative for classification tasks, by incorporating high level musical information directly in the preprocessing. To do so, we explore a new representation based on the \emph{rhythmic tree} structure, built from a context-free grammar tailored to symbolic music. We show that this representation outperforms existing encodings, such as tokenizations or piano rolls, on a drumming style classification built upon the GrooveMIDI Dataset \cite{groove2019}. In addition, our rhythmic tree-based encoding results in smaller deep models, with less parameters, able to be trained on less data compared to existing representations.

\begin{figure}[t]
     \centering
     \begin{subfigure}[t]{0.22\textwidth}
         \centering
         \includegraphics[alt={Image of a drum score}, width=\textwidth]{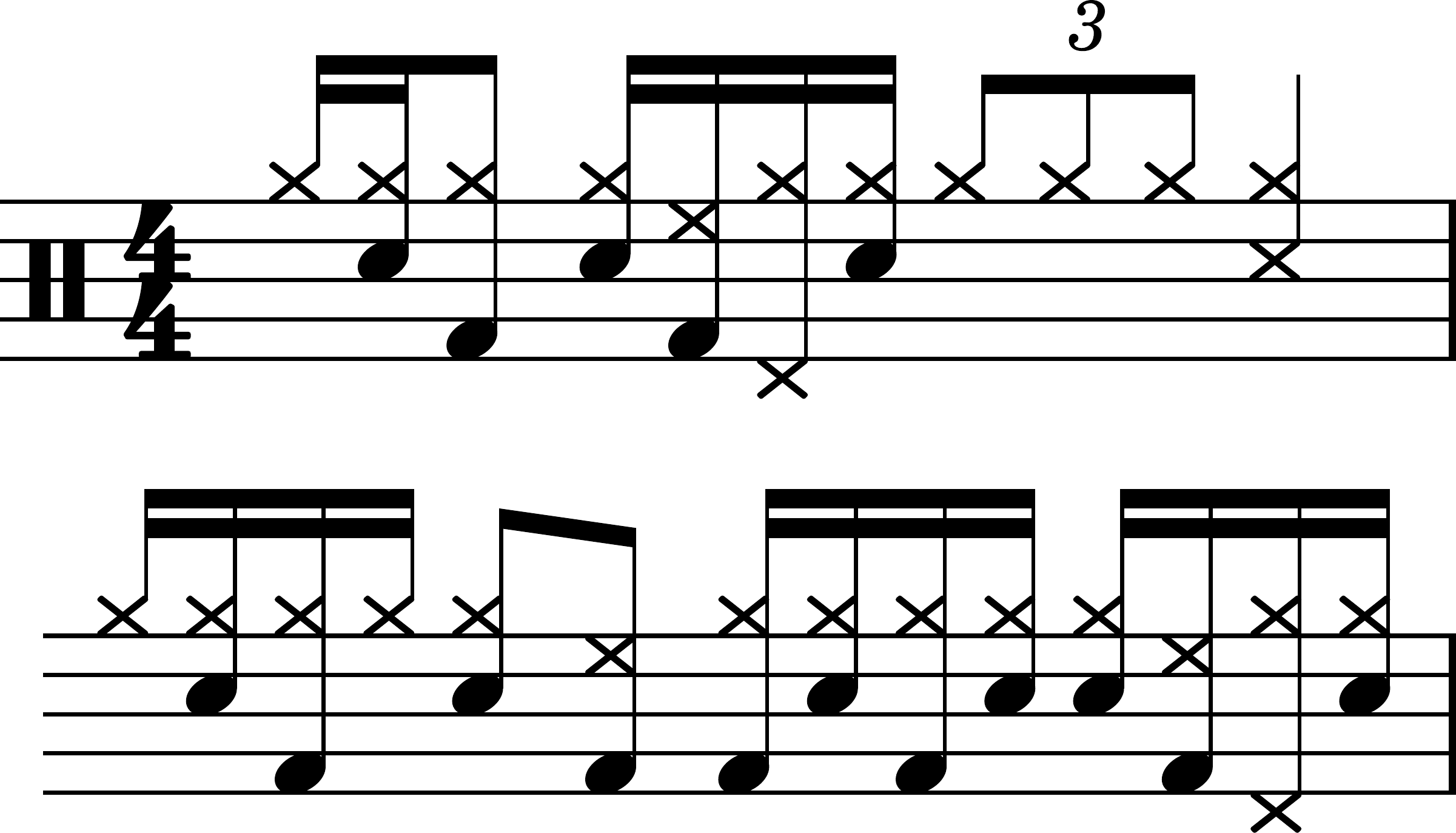}
         \caption{Base score}
         \label{fig:score}
     \end{subfigure}
     \begin{subfigure}[t]{0.25\textwidth}
         \centering
         \includegraphics[alt={Pianoroll of the drum score}, width=\textwidth]{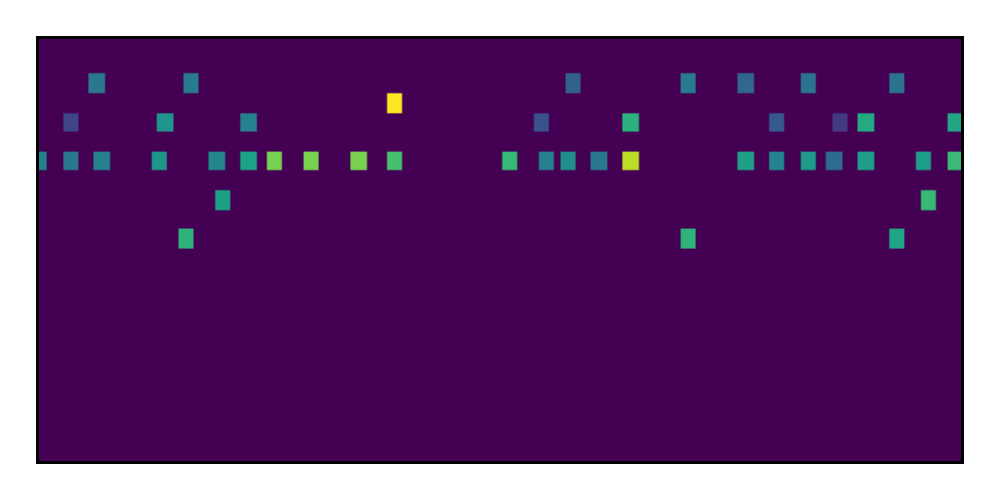}
         \caption{Piano roll}
         \label{fig:pianoroll}
     \end{subfigure}
     \begin{subfigure}[b]{0.22\textwidth}
        
\scriptsize\texttt{DrumOn\_42, Velocity\_43, TimeShift\_0.1.8, DrumOff\_42, TimeShift\_0.1.8, DrumOn\_42, Velocity\_39, DrumOn\_38, Velocity\_23, TimeShift\_0.1.8, DrumOff\_42, DrumOff\_38\dots}
        
         \caption{Tokenization (MIDI-Like)}
         \label{fig:midilike}
     \end{subfigure}
     \begin{subfigure}[b]{0.25\textwidth}
\hspace{-1.6mm}\scriptsize\texttt{
(p=42,v=0.35,d=0.1,t=0.000), 
(p=42,v=0.32,d=0.1,t=0.175), 
(p=38,v=0.17,d=0.1,t=0.006), 
(p=36,v=0.32,d=0.1,t=0.181), 
(p=42,v=0.35,d=0.1,t=0.017), 
(p=42,v=0.48,d=0.1,t=0.358), 
(p=38,v=0.41,d=0.1,t=0.034), 
(p=46,v=0.51,d=0.1,t=0.158)\dots}
         \caption{Note Tuples}
         \label{fig:note_types}
     \end{subfigure}
     \begin{subfigure}[b]{0.45\textwidth}
         \centering
         \includegraphics[alt={Linearized Rhythmic Tree matrix representation}, width=\textwidth,trim=0 10pt 0 0]{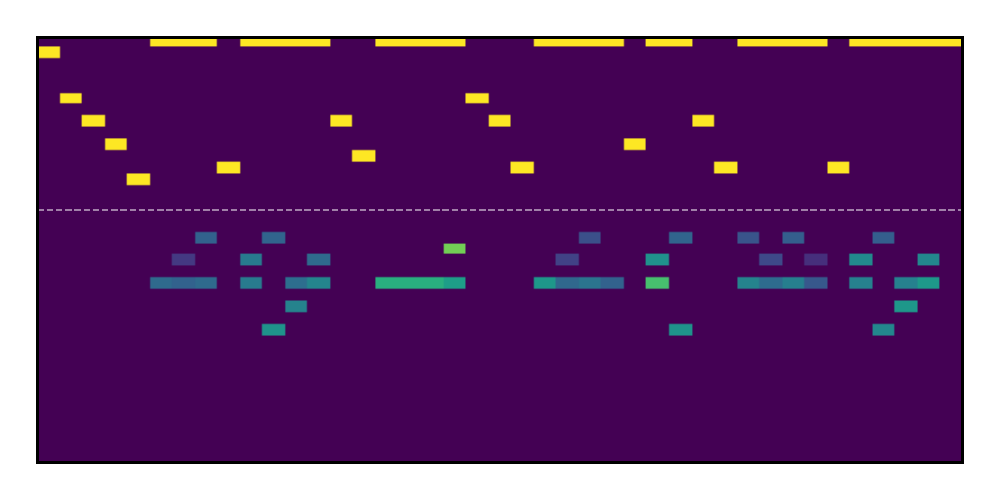}
         \caption{Linearized Rhythmic Tree}
         \label{fig:grammar}
     \end{subfigure}
        \caption{Different representations of the same two bars of drums. Score (\subref{fig:score}) is present for reference only.}
        \label{fig:representations_overview}
\end{figure}

\section{Background}\label{sec:background}

\subsection{MIDI Representations}\label{subsec:midi_representations}

MIDI is a lightweight musical information exchange format.
It does not carry audio data, but only timestamped events, \eg a note being played, featuring  its pitch and velocity, a note being released, a pedal change, etc.
It is suitable for recording as it captures the performer's expressiveness, but does not require metadata that are found in a score, such as tempo, time-signature\footnote{MIDI recordings can contain tempo and time-signature, but only through manual addition \textit{a posteriori}.}, tonality and voices \cite{wigginsFrameworkEvaluationMusic1993}. We discuss below the most common MIDI representations for ML.

\subsubsection{Piano Roll}

The piano roll is a visual representation of MIDI files  inspired by the analog rolls for piano players.
It consists in a 2D matrix with one dimension for pitches, and one for time. A note at pitch $p$ with a \verb|NOTE_ON| event at $x_\text{on}$ and \verb|NOTE_OFF| event at $x_\text{off}$ is given a positive value at positions $(x,p)_{x \in [x_\text{on}, x_\text{off}]}$, as shown in \cref{fig:pianoroll}. Often, the value in a matrix cell
is one of the properties of the MIDI event, \eg the velocity. 
This representation is popular, as its 2D structure allows to easily adapt deep models inspired by image processing (\eg CNN) to music tasks\cite{wangPerformanceNetScoretoAudioMusic2019,foscarinConceptBasedTechniquesMusicologistfriendly2022,velardeComposerRecognitionBased2016a}.
However, it can result in large sparse matrices with many zeros, since the time dimension must be discretized with a time step smaller than the shortest MIDI event.
In addition, piano rolls tend to be very long and redundant, since many successive vectors will be identical.

\subsubsection{Sequence of Tokens or Notes}

Similar to Natural Language Processing (NLP) techniques, recent works have adopted sequence-like representations, especially suitable for RNN and Transformers architectures. They encode MIDI files as sequences of events. These events are in turn transformed into \emph{tokens}, \ie discrete values from a vocabulary $V$. 
Many tokenizations exist, some consisting in a simple token/event mapping with MIDI files (MIDI-Like \cite{huangMusicTransformerGenerating2018,ooreThisTimeFeeling2020}, see \cref{fig:midilike}), while others include note durations (Structured \cite{hadjeresPianoInpaintingApplication2021a}, TSD \cite{bpe-symbolic-music}). More sophisticated tokenizers include higher level information about bar and position in the bar, such as REMI \cite{huangPopMusicTransformer2020a}.

Finally, MIDI files can be represented as ``note tuples'', \ie sequences of notes with attributes. For example, \cite{hawthorne2018transformer} represents each note by a set of four values: pitch, velocity, duration and time-shift compared to the previous note (cf. \cref{fig:note_types}). This representation is much more compact than piano rolls or sequence of tokens.

\subsection{Formal Grammar}

This work designs a symbolic music representation for deep networks based on a grammar-based rhythmic tree.
As a starting point, a formal grammar defines the syntax of a language $L$.
It consists in a set of symbols, associated with production rules used to rewrite non-terminal symbols into other (non-)terminal symbols.
Applied successively, those rules can produce every possible sentence of $L$.

\subsubsection{Context-Free Grammar}

Succinctly, a context-free grammar \cite{Hopcroft1979IntroductionTA} is a type of formal grammar for which the production rules do not depend on other context than the left-hand-side symbol.
It is defined as a 4-tuple $G = (V, \Sigma, R, S)$. $V$ is a finite set of non-terminal symbols, including the special \textit{start symbol} $S$. $\Sigma$ is a finite set of terminal symbols, called the alphabet. Finally, $R$ is a finite set of production rules of the form $a \rightarrow b$, where $(a, b) \in V \times (V \cup \Sigma)^*$ in which $^*$ denotes the Kleene star operator, \ie a pattern repeated of 0, 1  or more times.

The application of a sequence of rules can be represented as a tree, in which the parent node is represented by the left-hand-side of each rule, and the child nodes are the symbols on the right-hand-side.
Once every non-terminal symbol has been resolved into a terminal symbol, we obtain a \textit{parse tree} representing the structure of a sentence of $L$ according to $G$, with elements of $\Sigma$ as leaves, and $S$ as root.

\begin{figure*}[ht]
     \centering
     \includegraphics[alt={Example of the construction of a LRT, from the MIDI file to the final matrix}, width=\textwidth]{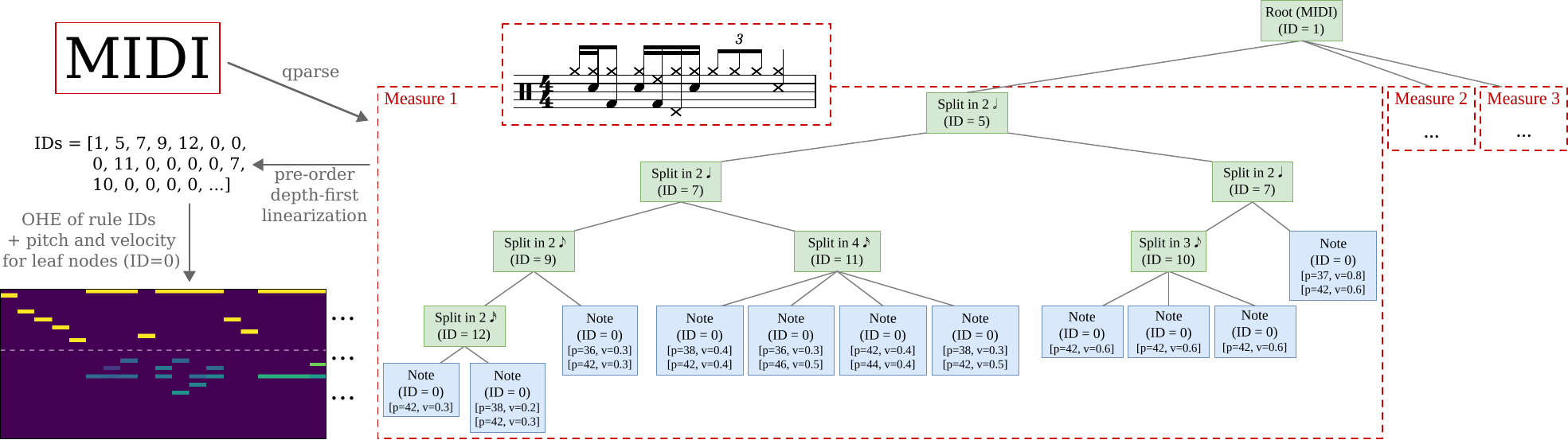}
     \caption{Example of tree built by \texttt{qparse} after rules simplification and re-rooting of measures (right), with its associated linearization and vector representation (left). In the matrix, the part above the dashed line contains the one-hot encoded rules (\textcolor{blue!50!black}{blue}/\textcolor{orange!70!black}{yellow} for 0/1), and the one below contains the playing instruments for terminal nodes (color representing velocity).}
     \label{fig:tree}
\end{figure*}

\subsubsection{Musical Grammar}\label{subsec:musical_grammar}

In a homophonic musical score (monophonic voice that can include chords \cite{amagasuTokenizationMIDISequences2024}), rhythm can be represented as a tree \cite{agonRepresentationRenderingRhythm2002,jacquemardStructuralTheoryRhythm2015}.
For example, in a 4/4 music piece, a measure could be split into two half notes. Then, each half note can be further divided into two quarter notes, or into a triplet of quarter notes, etc.
The \texttt{qparse} library \cite{foscarinParsebasedFrameworkCoupled2019b} is a MIDI-to-score transcription framework that produces a sheet music by parsing a MIDI file with a weighted context-free grammar and dynamic programming, with applications \eg to automatic drum transcription\cite{digardAutomatedTranscriptionElectronic2022a}.
While designed for a handcrafted music transcription algorithm, the intermediate parsing tree computed by \texttt{qparse} contains rich rhythmic information that is also valuable as an input to deep models.
Note that, while we our work uses \texttt{qparse} to obtain rhythmic trees, our contribution lies in evaluating this tree representation of music, regardless of its construction. We expect our representation to generalize to other parsers.

\section{Methodology}\label{sec:methodology}

\subsection{Linearized Rhythmic Tree}
\label{subsec:linear_grammar}

To build our high-level MIDI representation, we linearize a rhythmic tree obtained using a context-free grammar, enriched by information about pitch and velocity in leaves.
We call this representation Linearized Rhythmic Tree (LRT).
To achieve this goal, we leverage the transcription framework \texttt{qparse} \cite{foscarinParsebasedFrameworkCoupled2019b} to extract its internal intermediate rhythmic tree representation.
Note that we only consider homophonic inputs since this is what \texttt{qparse} MIDI grammar supports.
\texttt{qparse} needs the time-signature and the tempo of the track (because measures are parsed separately), as well as the specification of a weighted grammar.
We use a rhythm-oriented grammar similar to \cite{digardAutomatedTranscriptionElectronic2022a}, detailed in appendix.

As described in \cite{jacquemardSymbolicWeightedLanguage2022a}, the root of the intermediate rhythmic tree is the first measure. Its left child is a tree describing its beat decomposition, and its right child is a node pointing at the root of the next measure. We rewrite this tree so that all measures are children of the same global root. A $n$-measures-long track will therefore have a root with $n$ children. This rewriting allows us to reduce the maximum depth of the rhythmic tree, which would otherwise grow linearly with $n$. The resulting tree is shown on the right of \cref{fig:tree}.
In this tree, each node is labeled by the identifier of the associated production rule in the grammar.\footnotemark
Each leaf is a terminal symbol,  labeled by the note and properties from the associated MIDI events, \ie pitch\footnotetext{See the ruleset with IDs in Section 2 of the supplementary material.}\footnote{In the case of drums, the ``pitch'' corresponds to the drum used, \eg cymbal, snare, tom, etc.} and velocity.
Note that multiple instruments can be playing at the same time, so a leaf can be associated to several events.

As an example, in \cref{fig:tree}, the first bar (red frame) is split in two sections, each of half note length (rule 5). Then, the first half gets split into two quarter-length sections (rule 7). The second child of this node, a quarter note, is split into four sixteenth notes (rule 11). Finally, each of those sixteenth notes leads to a terminal symbol (rule 0), with MIDI events attached to it, \eg, the second child has two \verb|NOTE_ON| events, respectively with pitch 36 and velocity 0.3, and with pitch 46 and velocity 0.5.

As we cannot directly feed the tree structure to the models, we first linearize it using a pre-order depth-first traversal: we start from root, and traverse the nodes recursively following the left-most child, only going back up when the current branch has been fully traversed.
This produces a sequence of nodes containing the identifier of their rule in the grammar, as well as,
in the case of leaves, the list of playing instruments and their velocity.
We encode every node into a $d = (m + n)$-dimensional vector. $m$ is the number of rules in the grammar, and the first part of the vector is the one-hot-encoded identifier of the production rule associated with the node. $n$ is the number of possible instruments, and the second part of the vector contains the normalized velocity for each instrument. If an instrument is not playing for this note, its velocity is set to zero. For non-terminal rules, this second part is entirely zero. This linearization results in the matrix on the left of \cref{fig:tree}, \ie a sequence \mbox{$S = \{s\}_{t \in  \llbracket1,T\rrbracket}$} where $s_t \in \mathbb{R}^d$ is the vector associated to a node, and $T$ is the total number of nodes.
Therefore, our linearized rhythmic tree results in a multidimensional sequence $S$, that can be fed in all usual deep models such as RNN and Transformers.

Note that this representation is significantly shorter than tokenizations or piano rolls.
In average, the sequences are only around 18\% longer than note tuples, while containing much more information about the rhythm structure.

\subsection{Tree-based Positional Encoding for Transformers}\label{subsec:position_encoding}

While RNN can model the position in the sequence through their hidden state, Transformers process sequences as a bag of words, without any positional information.
To overcome this issue, positional encoding \cite{vaswaniAttentionAllYou2017} was introduced to incorporate information about the position of an element in the Transformer model. 

\newcommand\PE{\operatorname{PE}}
Classical positional encoding \cite{vaswaniAttentionAllYou2017} creates a vector $\PE$ of dimension $d$ using sine and cosine functions of increasing frequencies:
\newcommand\pos{\mathit{pos}}
\begin{equation}
    \omega_{\pos, i} = \frac{\pos}{\tau^{\left(\frac{2i}{d}\right)}},
    \begin{array}{l}
    \PE(pos,2i) = \sin\left( \omega_{\pos, i} \right)\\
    \PE(pos,2i+1) = \cos\left( \omega_{\pos, i} \right)
    \end{array}
    \label{eqn:pe}
\end{equation}
where $\pos$ is the position of the element in the sequence, $d$ the size of the embedding, $i \in \llbracket 1, d/2\rrbracket$ the dimension, and $\tau=10000$ as in \cite{vaswaniAttentionAllYou2017}.

\subsubsection{Continuous Positional Encoding}

For musical data, this positional encoding is not related to the temporal organization of the notes. Depending on how the sequence $S$ was built, the position $pos$ of an element can be arbitrary, such as \eg tokenizations where a note is split into several tokens for pitch, velocity and duration, or note tuples where two simultaneous notes can be interchanged.
For encoding note tuples, we therefore introduce a continuous positional encoding that replaces the position in the sequence by the timestamp of the note in the track: 
\begin{equation}
    \omega_{t,i} = \frac{2\pi}{T_S} \cdot \frac{t}{ \left( T_L/T_S \right) ^{\frac{2i}{d}}}, 
    \begin{array}{l}
    \PE(t,2i) = \sin\left( \omega_{t,i}\right)\\
    \PE(t,2i+1) = \cos\left(\omega_{t,i}\right)
    \end{array}
   \label{eqn:pe_cont}
\end{equation}
where $t$ is the absolute starting time of the note in seconds and $T_S$ and $T_L$ are respectively the smallest and largest periods of the sine functions.
This encoding allows two simultaneous notes to share the same positional encoding.

\subsubsection{Tree-based Positional Encoding}\label{subsubsec:hierarchical_pe}

A downside of linearizing the rhythmic tree is that we lose the explicit hierarchical structure between a parent node and its children.
The structure is still implicitly encoded in the linearized sequence $S$ in the rule identifiers, but the model would have to learn how the grammatical rules operate to rebuild the tree and leverage its structure.

\begin{figure}
     \centering
     \includegraphics[alt={Example of our tree-based positional encoding}, width=\linewidth]{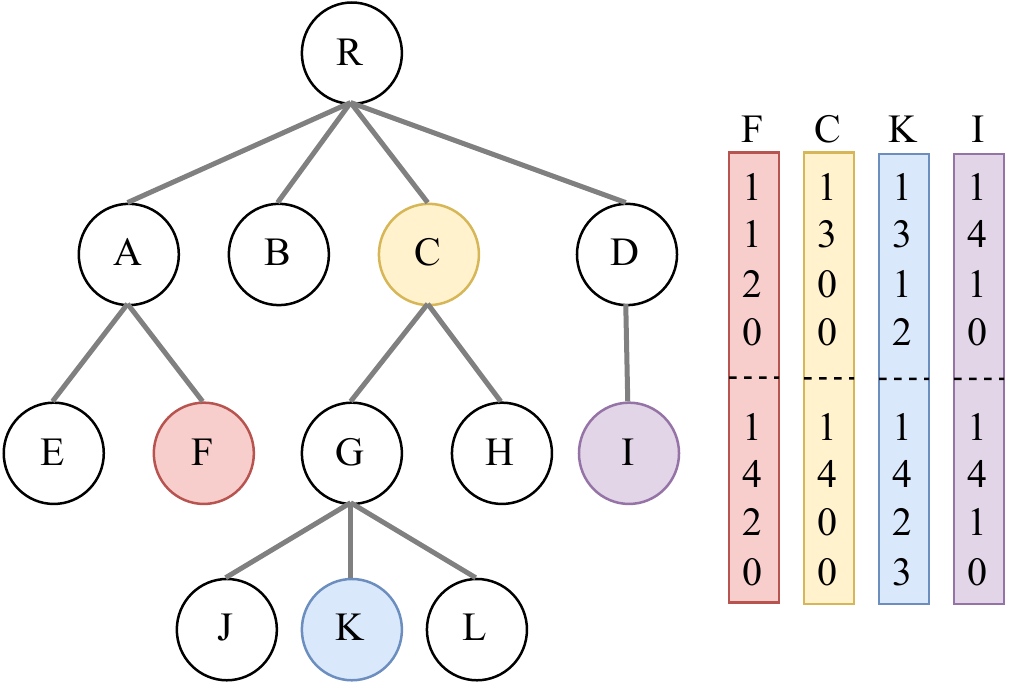}
     \caption{Example of tree-based positional encoding (TBPE) for a tree of maximum depth $d_\text{max}=4$.}
     \label{fig:tbpe}
\end{figure}

To better represent the rhythmic tree, we use a hierarchical tree-based positional encoding (TBPE) that encodes the position of a node \emph{in the tree}, rather than its position in the linearized sequence.
Some TBPE have been proposed in the literature, \eg for code translation to help Transformers process abstract syntax trees \cite{shivNovelPositionalEncodings2019a,pengRethinkingPositionalEncoding2022a}.
Since our trees are bounded in depth at $d_\text{max}$, we associate to each node $\mathcal{N}$ a vector of size $2 d_\text{max}$ that represents the path to a node from the root of the tree.
This process is illustrated in \cref{fig:tbpe}.
Element $k$ represents the index of the child traversed at depth $k$, while element $k + d_\text{max}$ is the total 
number of children of the parent node at depth $k$.
For example, to reach node $F$, we go through node $R$ (child \#1 over 1), then node $A$ (child \#1 over 4), then node $F$ (child \#2 over 2).
If the depth of the node $\mathcal{N}$ is less than $d_\text{max}$, then the remaining elements of the vector are padded with zeros.
This makes explicit in the positional encoding the \textit{parent $\rightarrow$ child}
relations, along with depth and breadth properties. It becomes easier for the model to understand that notes can belong to a larger structures (\eg triplet or four semiquavers).

\begin{table*}
\centering
\setlength\tabcolsep{3pt}

\begin{tabularx}{\textwidth}{ccc|cYccYc}
\toprule
\multicolumn{3}{c|}{\textbf{Representation used}} & \multicolumn{3}{c}{\textbf{LSTM}} & \multicolumn{3}{c}{\textbf{Transformer}} \\ \midrule
\multicolumn{1}{c|}{\textbf{Type}} & \multicolumn{1}{c|}{\textbf{Variation}} & \textbf{Avg. len.} & \textbf{Test F1 score} & \textbf{\# params.} & \textbf{\# bars} & \textbf{Test F1 score} & \textbf{\# params.} & \textbf{\# bars} \\ \midrule
\multicolumn{1}{c|}{\multirow{2}{*}{\textbf{Piano roll}}} & \multicolumn{1}{c|}{50 steps/second} & 4092 & \textit{0.618 $\pm$ 0.033} & \num{576522} & 4 & 0.545 $\pm$ 0.011 & \num{253130} & 2 \\
\multicolumn{1}{c|}{} & \multicolumn{1}{c|}{30 steps/second} & 2455 & \textbf{0.663 $\pm$ 0.023} & \num{552458} & 4 & 0.486 $\pm$ 0.026 & \num{20330} & 4 \\ \midrule
\multicolumn{1}{c|}{\textbf{Note Tuple}} & \multicolumn{1}{c|}{-} & 733 & 0.568 $\pm$ 0.024 & \num{555530} & 8 & 0.492 $\pm$ 0.014 & \num{216506} & 2 \\ \midrule
\multicolumn{1}{c|}{\multirow{4}{*}{\textbf{Tokenization}}} & \multicolumn{1}{c|}{MIDI-Like} & 2767 & 0.565 $\pm$ 0.026 & \num{42442} & 4 & 0.576 $\pm$ 0.011 & \num{360586} & 8 \\
\multicolumn{1}{c|}{} & \multicolumn{1}{c|}{REMI} & 2502 & 0.475 $\pm$ 0.051 & \num{38282} & 8 & 0.517 $\pm$ 0.028 & \num{17626} & 4 \\
\multicolumn{1}{c|}{} & \multicolumn{1}{c|}{Structured} & 2646 & 0.599 $\pm$ 0.014 & \num{282826} & 2 & \textit{0.598 $\pm$ 0.011} & \num{232162} & 8 \\
\multicolumn{1}{c|}{} & \multicolumn{1}{c|}{TSD} & 2464 & 0.487 $\pm$ 0.029 & \num{23274} & 8 & 0.486 $\pm$ 0.032 & \num{20178} & 4 \\ \midrule
\multicolumn{1}{c|}{\multirow{2}{*}{\textbf{LRT}}} & \multicolumn{1}{c|}{Simple linearization} & 863 & 0.603 $\pm$ 0.014 & \num{1358346} & 8 & 0.556 $\pm$ 0.037 & \num{230378} & 8 \\
\multicolumn{1}{c|}{} & \multicolumn{1}{c|}{With TBPE} & 863 & 0.596 $\pm$ 0.014 & \num{252170} & 4 & \textbf{0.660 $\pm$ 0.019} & \num{88138} & 4 \\ \bottomrule
\end{tabularx}

\caption{Performance of the different representations and model combinations on the GrooveMIDI dataset. We report macro F1 scores on the test set for the best model of each couple model/representation, alongside the model's number of parameters, the length (in bars) of input samples, and the average sequence length of each representation. Best results for each model type are in \textbf{bold}, second best in \textit{italics}.}
\label{tab:scores}
\end{table*}

\section{Experiments and results}

\subsection{Dataset and Task}

Our models are trained and evaluated for style classification on the Groove MIDI Dataset (GMD) \cite{groove2019}.
It consists in 13.6 hours of drumming music, played by humans with a metronome.
Each track is labelled with a style provided by the drummer, alongside tempo and time-signature.
The dataset is composed of long sequences (few minutes) and short beats and fills.
We only consider long sequences, as short sequences are less representative of a specific style. We also discard non-4/4 tracks (around 1\% of the dataset), as we use a 4/4 musical grammar, and a few tracks that \texttt{qparse} failed to parse\footnote{As these tracks are only in the train and validation sets, this does not affect the fairness of the final comparison.}.
We focus on the 4 most represented styles: funk, jazz, latin and rock.
The final subset contains 326 tracks, representing 7.5 hours of drumming, split into the train/validation/test sets (80\%/10\%/10\%) as the original dataset \cite{groove2019}. Each track is then further divided into multiple chunks of $n$ measures with a sliding window.

\subsection{Representations}\label{subsec:representations_evaluated}

In addition to our LRT, we evaluate common representations of MIDI data for style classification.

\textbf{Piano Roll}
We sample the MIDI data at frequency $f$. We compare $f = \SI{30}\hertz \approx \qty{33.3}{\milli\second}$ per time step, as \SI{30}{\milli\second} is considered as the simultaneity threshold for the human ear \cite{goeblMelodyLeadPiano2001}, and $f = \SI{50}\hertz = \qty{20}{\milli\second}$ per time step, to see if models would improve with finer granularity, at the expense of sequence length.
Every time step is represented by a vector in $v \in [0,1]^{22}$. Each dimension represents one of the 22 instruments of the drum kit.
$v_i$ encodes the velocity of the $i$-{th} instrument, normalized between 0 and 1 using maximum normalization.
Note that the duration of notes in drums MIDI files is arbitrary, as only onset and velocity matter. All durations are set to \SI{100}{\milli\second} in the Groove MIDI dataset.
In our dataset, the average length of a piano roll is around 2455 for $f = \SI{30}\hertz$, and 4092 for $f = \SI{50}\hertz$.

\textbf{Sequence of Tokens}
We experiment with various tokenizers from the literature, that quantify velocities and timings to limit the size of the vocabulary: MIDI-Like \cite{huangMusicTransformerGenerating2018,ooreThisTimeFeeling2020}, TSD \cite{bpe-symbolic-music}, Structured \cite{hadjeresPianoInpaintingApplication2021a} and REMI \cite{huangPopMusicTransformer2020a} tokenizers.
We use the default parameters from \cite{miditok2021}, except for pitch range which is set to the min/max instrument ID from the GMD.
Models trained on tokenizations use a 64-dimensional embedding, as recommended in\cite{zhangSymbolicMusicRepresentations2023b}.
Akin to piano rolls, tokenizers produce sequences with 2400 to 2800 elements.

\textbf{Note Tuples}
We also consider the note tuples \cite{hawthorne2018transformer} representation that uses a single vector for each note.  Each vector has 25 dimensions: the 22 one-hot-encoded instrument, followed by normalized velocity, note duration and time-shift to the previous note.
This results in shorter sequences, with as many elements as there are notes. 
Average sequence has 733 elements, 3.5$\times$ less than tokenization methods.

\textbf{Linearized Rhythmic Tree}
We use a simplified rhythm grammar of 15 rules on the GMD. As this grammar does not allow notes shorter than a 1/32nd note, the maximum depth $d_\text{max}$ of a leaf in the rhythmic tree is 6.
Although slightly longer than note tuples, the resulting sequences remain on the smaller side with an average of 863 elements.

\subsection{Models}

We chose to focus on sequential representations and therefore consider two popular architectures: LSTM \cite{hochreiterLongShorttermMemory1997} and Transformers \cite{vaswaniAttentionAllYou2017}.
The model inputs are fed as chunks of 2, 4 or 8 measures.
We perform a hyperparameter search for the number of bars, number of layers and layer width on the validation set and retain the best architectures for each (model, representation) combination.
As our grammar parser uses the track's tempo, we inject this information in non-grammatical models for a fair comparison by concatenating the tempo to the features vector in the last layer.

\textbf{LSTM architecture}
We consider  bidirectional LSTM models \cite{gravesFramewisePhonemeClassification2005} and we experiment with a depth of 1 to 4 layers and a fixed width of 8 to 256 neurons per layer.
Even though LSTMs do not require positional encoding, we also evaluate our LRT representation with TBPE to assess whether the explicit rhythmic structure is beneficial to the model.

\textbf{Transformer architecture}
We use standard Transformers with an embedding layer, \ie a linear projection, between the input and the first Transformer block. The models have 1 to 4 encoder layers, each with 2 to 16 attention heads. We also experiment with a feature size of 2 to 32 dimensions per head and 8 to 64 neurons in the feedforward network. We use the classical positional encoding for token sequences, the continuous positional encoding for piano rolls and note tuples, and either the classical or the tree-based positional encoding for LRTs.
Regarding continuous encoding, we use $T_S = \SI{100}{\milli\second}$ so that even close notes have a different encoding, and $T_L = \SI{300}{s}$, as temporal context is unlikely to matter beyond several minutes.

Final models are trained with a batch size of 128, using the AdamW optimizer \cite{loshchilovDecoupledWeightDecay2018} with a learning rate of 0.001, decayed by a factor 10 every 50 epochs with weight decay and dropout.
Early stopping occurs when the validation F1 score plateaus with a patience of 200 epochs.
Models are trained using the standard cross-entropy loss. To alleviate the class imbalance (185 rock tracks versus 50 for the other classes), we use class inverse median frequency weighing.
We report the macro F1 scores
averaged over all classes.

\subsection{Main Results}

We report in \cref{tab:scores} the test scores of the best combinations from the hyperparameter search, averaged over five runs.

LSTM with $\qty{30}{\hertz}$ piano rolls and Transformer with LRT/TBPE are the combinations that lead to the best F1 scores overall ($\approx 0.66$).
The former is a 3-layer LSTM model, each composed of 64 neurons, performing on 2-bar-long samples. The latter is a 4-layer Transformer model, each using 2 heads with 32 features per head (so a 64-dimensional input vector), and a feedforward network of 32 neurons trained on 4-bar-long chunks.
Although both models achieve comparable performance, note that the Transformer model needs $6\times$ fewer parameters than the LSTM.

We observe that the TBPE provides important information for style classification. Transformer models using a classical positional encoding achieve lower classification performance ($\approx 0.56$). Surprisingly, using TBPE is beneficial for LSTMs also: both our LSTM models trained on LRT achieve nearly identical F1 scores ($\approx 0.6$), however injecting the TBPE allows us to use a RNN with $5\times$ fewer parameters. This confirms that explicitly encoding the node position in the tree makes it easier for the models to understand the rhythmic structure of the track.

Finally, we observe that tokenization and note tuples tend to underperform overall.
Structured MIDI tokenization achieves the best of tokenizer F1 score ($\approx 0.6$) both for LSTM and Transformer architecture, followed by MIDI-Like, however at the cost of a higher number of parameters.
Token or note tuple sequences seem difficult to learn for the models.
For RNN, we hypothesize that this is due to the regular sampling assumption made by these models. Each element is processed by the same recurrent loop, meaning that the model needs to learn the structure of the sequence, \eg what each token represents. In comparison, piano rolls with a fixed time step where all elements represent the same object tend to have higher performances with LSTMs.

\subsection{Model Parameter Efficiency}\label{subsec:num_params}

\begin{figure}
     \centering
     \includegraphics[alt={Plot of the validation F1-scores of variations of our best models depending on the number of parameters of the model}, width=\linewidth]{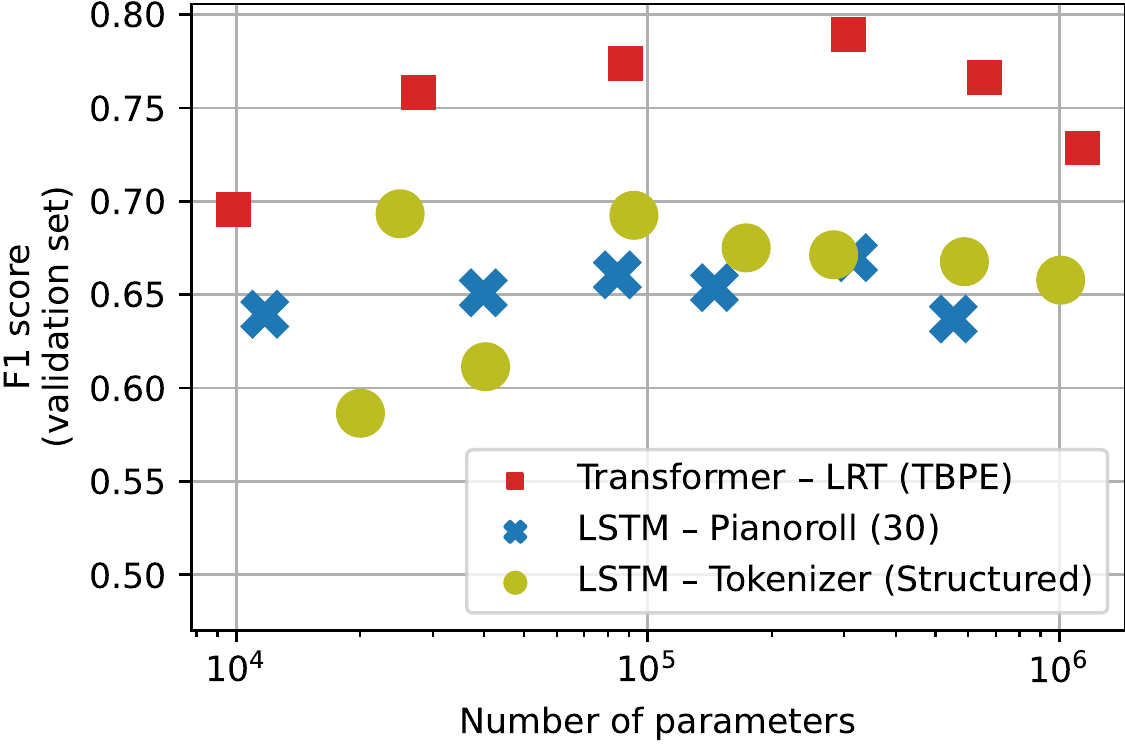}
     \caption{F1 scores on the validation set vs. number of parameters for a selected set of models. We observe that Transformers trained on LRT consistently outperform other models at similar capacity.}
     \label{fig:num_params}
\end{figure}

We evaluate some representative models by varying their capacity, \ie number of parameters. More specifically, we experiment with 4, 8, 32 and 64 number of features per head for the Transformer, and 16, 32, 48, 64, 96 and 128 neurons in the hidden layers for LSTM.
We report F1 scores on the validation set in \cref{fig:num_params}. We observe that, at comparable number of parameters, the Transformer trained on the LRT always lead to higher F1 scores than the compared models.
This demonstrates that the rhythmic information embedded in our rhythmic tree not only results in shorter sequences, but also can be leveraged by smaller models for better or
on par performance compared to existing works.

\subsection{Training Samples Efficiency}

\begin{figure}
     \centering
     \includegraphics[alt={Plot of the validation F1-scores of our best models depending on the dataset size}, width=\linewidth]{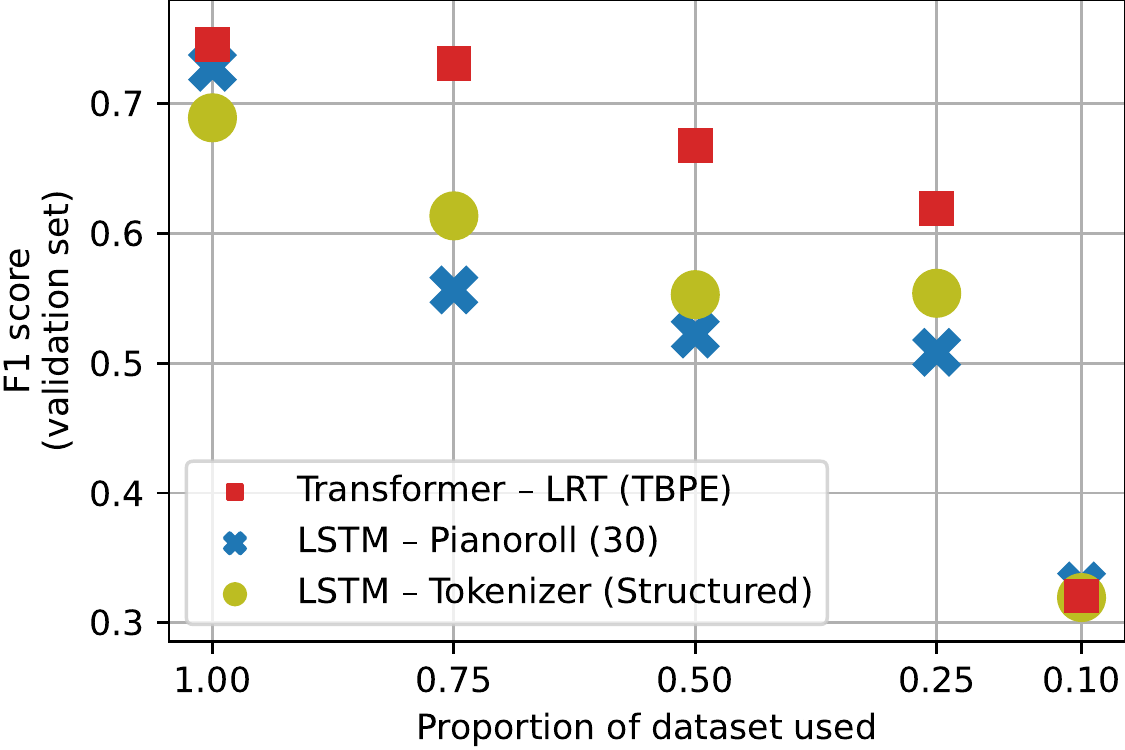}
     \caption{F1 scores on the validation set vs. percentage of training samples used. Transformers trained on LRT exhibit a less severe performance drop when the number of training samples decreases compared to existing models.}
     \label{fig:ablation}
\end{figure}

Finally, we evaluate how representation affects the amount of data needed to train our models. We compare the same models as in \cref{subsec:num_params} and train them with a random subset of 75\%, 50\%, 25\% and 10\% of the training set.
F1 scores on the validation set are reported in \cref{fig:ablation}.
We observe that the Transformer model trained with the linearized rhythmic tree and the tree-based positional encoding consistently outperforms the structured tokenizer and the piano roll. 
The performance drop between 100\% and 75\% is minimal, and overall the LRT-based Transformer degrades more gracefully when the number of training samples decreases compared to the other models.
This underlines the relevance of the LRT, that encodes higher level musical information and better represents the invariance of musical style to spurious variations in the input MIDI file, such as slight changes in timings or velocity.

\section{Conclusion and Future Work}

We evaluated different representations of MIDI data for drumming style classification.
We introduced a new representation based on the linearization of a rhythmic tree obtained by parsing a MIDI file using a musical grammar. This representation provides richer features while being more compact than traditional piano rolls or tokenization strategies.
Associated with a Transformer architecture using a tree-based positional encoding, we show that this representation achieves style classification performance on par with the best models from the literature with much fewer parameters.
We also provide evidence that our representation is more resilient when trained on smaller datasets.

Future works involve extending this tree-based representation beyond homophonic input, \eg for polyphonic piano pieces. Building the parsing tree could also be achieved on music scores, making it possible to directly classify scores at the mere symbolic level.
In addition, we would like to evaluate this approach on more diverse tasks, as representation could be beneficial not only for discriminative models, but also for generative models, \eg in music generation tasks, to produce syntactically correct performances with respect to the specified grammar\cite{kusnerGrammarVariationalAutoencoder2017b}.

\section{Acknowledgements}

We thank Florent Jacquemard for his work on \texttt{qparse}, fruitful discussions on the design of the rhythmic grammar for drums and advice throughout this project. Additional thanks are dedicated to Lydia Rodriguez de la Nava for her help in adapting \texttt{qparse} to drums rhythm parsing.
\bibliography{references}

\end{document}


\maketitle
\vspace{-2.5cm}

\section{Raw weighted grammar used with \texttt{qparse}}

The following file represents the grammar used by \texttt{qparse} to build the parsing tree. The rules represent the divisions of the time that are allowed during the parsing. Each rule has an associated weight for \texttt{qparse} to choose the simplest tree possible. For more details, please refer to the original \texttt{qparse} article and the documentation of the library.

\begin{lstlisting}
[penalty]

[timesig 4 4]

// bar level (whole)
0 -> C0          1
0 -> E1          1     // whole
0 -> E2          1.55  // flam + whole 
0 -> U2(1, 1)    0.1   // div. in 2 halves
0 -> U3(1, 1, 1) 33    // triplet of halves

// half bar level (half)
1 -> C0           1
1 -> E1           1      // half
1 -> E2           1.85   // flam + half
1 -> U2(2, 2)     0.1    // beats (quarters)
1 -> U3(2, 2, 2)  33     // triplet of quarters

// beat level (quarter)
2 -> C0             1     // tied quarter
2 -> E1             1     // quarter
2 -> E2             2.25  // flam + quarter
2 -> T2(3, 3)       0.1   // 2 eights
2 -> T3(3, 3, 3)    3.25  // triplet of eights    
2 -> T4(4, 4, 4, 4) 0.15  // 4 16th

// sub-beat (eight)
3 -> C0           1      // tied eighth
3 -> E1           1      // eight
3 -> E2           2.35   // flam + eighth
3 -> T2(4, 4)     0.9    // 2 16th

// sub-sub-beat (16th)
4 -> C0        1      // tied 16th
4 -> E1        1      // 16th
4 -> E2        4.15   // flam + 16th
4 -> T2(6, 6)  0.65   // 2 32d

// triple (32d)
6 -> C0        1      // tied 32d
6 -> E1        1      // 32d
\end{lstlisting}

Note that while this specific grammar can only parse triplets of half, quarter and eight notes, it could be easily modified to take into account other triplets (\eg triplets of sixteenth notes) or more diverse tuplets (\eg quintuplets, septuplets).

Also note that the parsing tree represents a rhythm, and not directly notes. Therefore, some branches of the parsing tree might not represent any played note: a dotted half note followed by a quarter note in a 4/4 measure will result in a tree with 2 branches, each of them subdivided into two branched. However, only the first and the last leaves represent played notes, the 2nd and 3rd leaves are only continuations of the dotted half note (first leaf).

\section{Simplified grammar used in our trees}

When building our trees, we simplified the tree built by \texttt{qparse} to use a subset of rules in order to get a simpler representation. We associated each rule with an identifier, and gathered together under a same ID (0) all terminal rules (rules leading to a terminal symbol like a note, a rest).

\subsection{Non-terminal symbols}

\begin{lstlisting}
S:   start symbol
n1:  whole-note-long segment
n2:  half-note-long segment
n4:  quarter-note-long segment
n8:  eighth-note-long segment
n16: sixteenth-note-long segment
n32: thirty-second-note-long segment
n*:  anyone of the previous segment symbols
\end{lstlisting}

\subsection{Terminal symbols}

\begin{lstlisting}
note: rhythmic element (note, rest, or continuation of the previous leaf)
\end{lstlisting}

\subsection{Production rules}

\begin{lstlisting}
 0 n* -> note                  // only rule leading to terminal symbol
 1 S  -> (n1, n1...)           // root of the tree, parent of all measures
 2 currently unused
 3 currently unused
 4 currently unused
 5 n1  -> (n2, n2)             // whole note into two half notes
 8 n2  -> (n4, n4, n4)         // whole note into triplet of half notes
 7 n2  -> (n4, n4)             // half note into two quarter notes
 8 n2  -> (n4, n4, n4)         // half note into triplet of quarter notes
 9 n4  -> (n8, n8)             // quarter note into two eighth notes
10 n4  -> (n8, n8, n8)         // quarter note into triplet of eight notes
11 n4  -> (n16, n16, n16, n16) // quarter note into four 16th notes
12 n8  -> (n16, n16)           // eighth note into two 16th notes
13 n16 -> (n32, n32)           // 16th note into two 32nd notes
14 currently unused
\end{lstlisting}

\section{Confusion matrix of best model}

\begin{figure}[H]
     \centering
     \includegraphics[alt={Confusion matrix of best model}, width=0.6\textwidth]{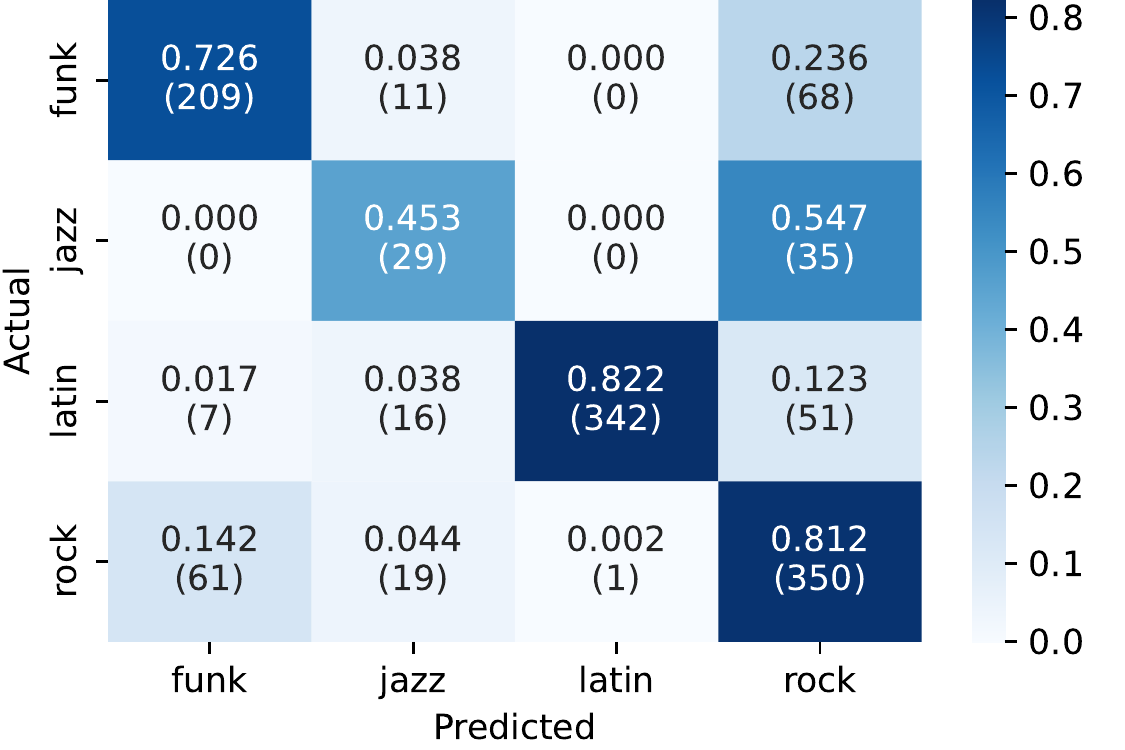}
     \label{fig:cm}
\end{figure}

This is the confusion matrix of our best model: transformer associated with TBPE. The matrix is reported for the validation set, and is normalized by row (total of samples with same true style).

As we could expect, the most difficult style to classify is the jazz. Indeed, jazz music can be very diverse, and it might be difficult for a model to extract features that are specifically tied to jazz. Note that jazz tracks are under-represented compared to the three other styles in the validation set, but that it is not the case in the training set.

We can also see that misclassified non-rock samples tend to be classified as rock. The reason is probably the over-representation of this style in the training set, despite our attempt to normalize it in the loss.

The confusion matrix of other models were generally quite similar to this one, revealing notable difficulties with jazz samples and more affinity for rock than other styles.